\documentclass[preprint,12pt]{elsarticle}

\usepackage{epsfig}
\usepackage{enumerate}

\usepackage{amssymb,amsmath,graphics,graphicx}  
\usepackage[dvipsnames]{xcolor}

\journal{Elsevier}

\begin{document}

\begin{frontmatter}


\title{Collective attention under digital exposure: A dynamical systems approach}

\author{Nuno Crokidakis}
\ead{nunocrokidakis@id.uff.br}

\address{Instituto de F\'{\i}sica, Universidade Federal Fluminense, Niter\'oi, Rio de Janeiro, Brazil}

\begin{abstract}
  The widespread use of digital devices has raised growing concerns about its impact on sustained attention at the population level. In this work, we propose a minimal dynamical framework to describe the collective evolution of attention under continuous exposure to screen-mediated environments. We introduce a macroscopic variable representing the population-level sustained attention and model its dynamics as the result of competing mechanisms: intrinsic cognitive recovery and degradation induced by digital stimulation. The digital environment is treated as an external control parameter that continuously perturbs the system, leading to a relaxational dynamics. The proposed mechanisms are consistent with empirical findings on attentional dynamics under digital exposure. We first analyze a linear formulation, which provides an analytically tractable baseline, and then extend the model by incorporating a nonlinear degradation term that captures amplification effects under high-intensity stimulation. We derive an explicit expression for the stationary state and show that the equilibrium attention level decreases monotonically with increasing exposure. An effective potential formulation is introduced, revealing that digital overstimulation progressively deforms the dynamical landscape, shifting the stable state toward regimes of reduced attention without generating multiple equilibria. Importantly, the model does not rely on social contagion or interaction-driven bistability, but instead describes a continuous displacement of the collective cognitive regime under environmental pressure. Our results suggest that the impact of digital technologies on attention may be understood as a gradual macroscopic effect emerging from persistent external stimulation, rather than as a transition between competing behavioral states.

\end{abstract}


\end{frontmatter}



\section{Introduction}

The rapid expansion of digital technologies has profoundly transformed the way individuals interact with information. In particular, the widespread use of smartphones and online platforms has led to a cognitive environment characterized by continuous stimulation, rapid information switching and short feedback cycles. While these technologies have brought undeniable benefits, growing empirical evidence suggests that they may also affect fundamental cognitive processes \cite{arruda,yakura}, especially sustained attention.

A number of studies in psychology and neuroscience have reported correlations between intensive screen use and reduced attention span, increased distractibility, and difficulties in maintaining focus over extended periods \cite{Twenge,mark1,mark2,mark3,gazzaley}. In parallel, behavioral studies indicate that digital environments promote frequent task switching and fragmented patterns of information consumption, which may further challenge the maintenance of sustained cognitive effort \cite{mark1,ophir,leroy}.

Despite these advances, most existing approaches focus on individual-level mechanisms, such as cognitive load, attentional capture or reward processing \cite{gazzaley,kahneman}. In many complex systems, macroscopic dynamical equations can be derived from explicit microscopic interactions through evolutionary, agent-based or game-theoretical frameworks \cite{Traulsen2005,Traulsen2006,Tu2023}. Such approaches often rely on coarse-graining or dimensionality-reduction procedures that connect high-dimensional microscopic dynamics to low-dimensional collective descriptions. The present work follows a complementary strategy by proposing a phenomenological coarse-grained framework formulated directly at the population level, without attempting an explicit derivation from an underlying networked system. However, comparatively little attention has been devoted to simple population-level dynamical frameworks aimed at describing how persistent digital exposure may shape collective attention. Such effects may emerge from the interaction between individual cognition and a shared technological ecosystem \cite{rmp}.

Recent works have proposed mathematical and computational approaches to describe attention dynamics in digital environments, particularly at the level of individual behavior and in the context of online social media activity \cite{vacaruiz,ojer}. These studies provide important insights into how attention is allocated, fragmented and depleted under competing stimuli and rapidly changing information streams. However, most existing models remain focused on microscopic or data-driven descriptions, often lacking a coarse-grained representation of attention at the population level. In particular, the formulation of minimal dynamical frameworks capable of capturing the collective evolution of sustained attention under persistent environmental exposure remains largely unexplored. In contrast, empirical findings discussed above, such as increased task switching, cognitive fragmentation, and reduced sustained focus in digital contexts \cite{Twenge,mark1,mark2,mark3,gazzaley}, suggest that attention may also be understood as a macroscopic variable shaped by shared external conditions.

From a theoretical perspective, this raises a natural question: can the large-scale effects of digital exposure on attention be understood within a simple dynamical framework, in analogy with other macroscopic phenomena studied in statistical physics?

Statistical physics has proven to be a powerful tool for describing collective behavior in social and cognitive systems. Models originally developed to study magnetic systems and phase transitions have been successfully adapted to problems such as opinion dynamics, cultural dissemination and collective decision-making \cite{rmp,galam}. In these contexts, the focus is typically on the emergence of collective states through interactions between agents, often leading to bistability or phase transitions between competing configurations.

However, the problem addressed in this work is conceptually different. Rather than modeling the spread of opinions or behaviors through social interaction, we consider the possibility that the digital environment acts as an external field that continuously modifies the intrinsic cognitive dynamics of the population. In this view, the key variable is not the fraction of individuals adopting a given state, but the population-level sustained attention as a macroscopic quantity.

This shift in perspective leads to a distinct modeling approach. Instead of interaction-driven dynamics \cite{racism}, we focus on the competition between two fundamental processes: (i) recovery mechanisms that restore cognitive resources, and (ii) degradation mechanisms associated with persistent exposure to digitally mediated stimuli. The latter is modeled as a continuous external perturbation, whose intensity controls the long-term behavior of the system.

The central objective of this paper is to propose and analyze a minimal dynamical model capturing this competition. The goal is to not provide a detailed description of cognitive processes, but rather to identify a minimal mechanism capable of reproducing generic trends observed at the population level. We show that, even in the absence of explicit social interactions, the system exhibits nontrivial collective behavior at the macroscopic level. In particular, the model predicts a continuous decrease in the stationary attention level as digital exposure increases. The present framework is not intended as a microscopic cognitive model, but rather as an effective coarse-grained description of population-level sustained attention.

To gain further insight, we introduce an effective potential formulation, which allows us to interpret the dynamics in terms of a deformable landscape. Within this framework, increasing exposure does not generate competing stable states, but rather continuously shifts the unique stable equilibrium toward lower levels of attention. This provides a physically transparent picture of how digital environments reshape collective cognitive regimes. This interpretation further emphasizes the role of external forcing in driving the system away from high-attention states.

The present work thus contributes to the growing interface between statistical physics and cognitive and social dynamics by introducing a framework in which environmental factors, rather than interactions, play the dominant role. More broadly, it suggests that some of the large-scale effects associated with digital technologies can be understood as emergent macroscopic consequences of persistent external stimulation.


\section{Models and Results}

\subsection{General model formulation}

We define a dynamical variable $x(t)$ representing the population-level sustained attention of the population, which takes values in the interval $[0,1]$. Here, $x=1$ corresponds to a regime of high sustained attention, while $x=0$ represents a strongly fragmented cognitive state. The variable $x(t)$ represents the collective level of sustained attention in the population and should be interpreted as a coarse-grained macroscopic descriptor rather than as the average of explicitly modeled individual cognitive states.

The digital environment is modeled as an external perturbation that continuously affects this cognitive capacity. This effect is captured by a control parameter $T \geq 0$, which represents the intensity of digital exposure. In a coarse-grained sense, $T$ may incorporate factors such as average screen time, density of digital stimuli, prevalence of fast-reward platforms and the overall cognitive pressure imposed by digital systems.

We assume that the interaction of individuals with the digital environment induces a degradation process that reduces the level of sustained attention. At the same time, we consider the existence of intrinsic recovery mechanisms, associated with offline or low-stimulation activities, which act to restore attention. This recovery process is characterized by a parameter $r$, representing the rate of cognitive restoration.

The dynamics of the system is thus governed by the competition between these two mechanisms, degradation and recovery, which together determine the collective attention level in the long-time limit. In this framework, $x(t)$ plays the role of a macroscopic variable evolving under the influence of an external field, in analogy with relaxational dynamics in driven systems \cite{Hohenberg,Goldenfeld}.

Although the present framework is formulated directly at the macroscopic level, one may interpret the proposed dynamics as an effective coarse-grained description emerging from the aggregate behavior of individuals subject to competing restorative and disruptive cognitive influences. In this perspective, deriving the macroscopic equations from explicit microscopic or agent-based attention dynamics would constitute an interesting direction for future investigation.

We first introduce a minimal linear model to establish the basic mechanism, and then extend it by incorporating nonlinear effects that capture additional features of cognitive degradation.


\subsection{Minimal linear model}

The simplest formulation of the model described in general terms in the previous subsection consists of assuming a linear form for the degradation process. Together with the recovery term, this leads to the following differential equation governing the evolution of $x(t)$:
\begin{equation} \label{eq1}
\frac{dx}{dt}=r(1-x)-\alpha\,T\,x ~.
\end{equation}
\noindent
Here $x$ and $\alpha$ are dimensionless quantities, whereas $r$ and $T$ have dimensions of inverse time. The first term in Eq. \eqref{eq1} represents the intrinsic tendency of attention to recover. This process may be associated with rest, sustained cognitive activities such as long-form reading, interactions outside digital environments and other habits that promote focus. The factor $(1-x)$ ensures that recovery is more effective when attention level is low, while it saturates as $x\to 1$.

The parameters of the model should be interpreted as effective coarse-grained quantities representing average population-level tendencies rather than fixed individual cognitive traits.

The second term encodes the degradation induced by digital exposure, i.e., the loss of sustained attention due to interaction with screen-mediated environments. This contribution is proportional to $x$, reflecting the fact that only the existing level of attention can be degraded. The parameter $\alpha>0$ quantifies the sensitivity of the population to the digital environment.

Equation \eqref{eq1} thus defines a minimal dynamical description of the competition between recovery and degradation processes, providing a baseline for further extensions.

We can also introduce dimensionless variables $\tilde{t}=r\,t$ and $\theta=\alpha\,T/r$. In terms of these variables, Eq. \eqref{eq1} becomes
\begin{equation} \label{eq1_2}
\frac{dx}{d\tilde{t}}=(1-x)-\theta\,x ~.
\end{equation}
\noindent
showing explicitly that the dynamics is governed by a single dimensionless control parameter, $\theta$, which quantifies the relative strength of digital exposure with respect to the intrinsic recovery rate.

Taking the long-time limit $t\to\infty$ in Eq. \eqref{eq1}, we obtain the stationary population-level sustained attention in the population,
\begin{equation}  \label{eq2}
x^{*} = \frac{r}{r + \alpha\,T}  ~.
\end{equation}
\noindent
Eq. \eqref{eq2} shows that $x^{*}=1$ when $T=0$, i.e., in the absence of digital exposure, the attention level reaches its maximum value. As $T$ increases, corresponding to higher levels of exposure, the stationary attention level $x^*$ decreases monotonically, approaching $x^*\to 0$ in the limit $T\to\infty$.

Eq. \eqref{eq1} can also be solved analytically for arbitrary times. Considering the initial condition $x(0)=x_0$, directed integration yields
\begin{equation}  \label{eq3}
x(t) = x^* + (x_0 - x^*)\,e^{-(r+\alpha\,T)t} ~.
\end{equation} 
\noindent
which shows that the stationary value $x^*$ is reached exponentially. From Eq. \eqref{eq3}, the characteristic relaxation time is given by $\tau=1/(r+\alpha\,T)$. Thus, the convergence toward the stationary state becomes faster as $T$ increases, while the final attention level is simultaneous reduced. In a similar way, the model's solution can also be rewritten in terms of $\tilde{t}$ and $\theta$, leading to
\begin{equation}   \label{eq3_2}
x(\tilde{t}) = x^* + (x_0 - x^*)\,e^{-\theta\,\tilde{t}} ~.
\end{equation}

Expressing the stationary solution in terms of the dimensionless parameter $\theta=\alpha\,T/r$, Eq. \eqref{eq2} can be rewritten as $x^*=1/(1+\theta)$. This expression shows that the dynamics is governed by the ratio between digital exposure $\alpha\,T$ and recovery capacity $r$. In the limit of large $T$, one obtains the asymptotic scaling $x^*\sim \theta^{-1}\sim T^{-1}$. 

We thus see that even this minimal linear formulation captures key qualitative features of the phenomenon. In particular, the collective level of sustained attention can be interpreted as a macroscopic dynamical variable governed by the competition between cognitive recovery and degradation induced by the digital environment.

\begin{figure}[t]
\begin{center}
\vspace{6mm}
\includegraphics[width=0.35\textwidth,angle=0]{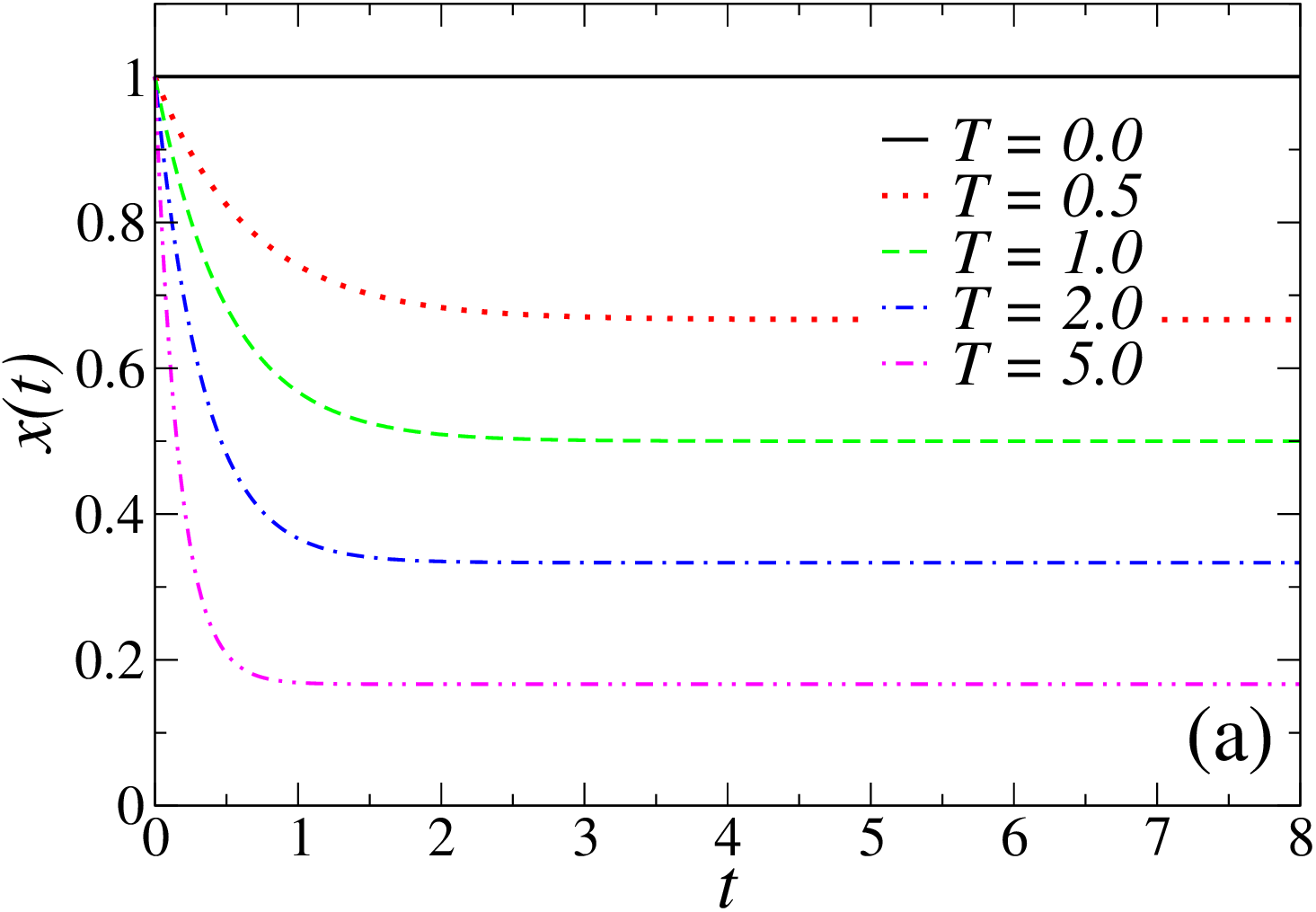}
\hspace{0.3cm}
\includegraphics[width=0.35\textwidth,angle=0]{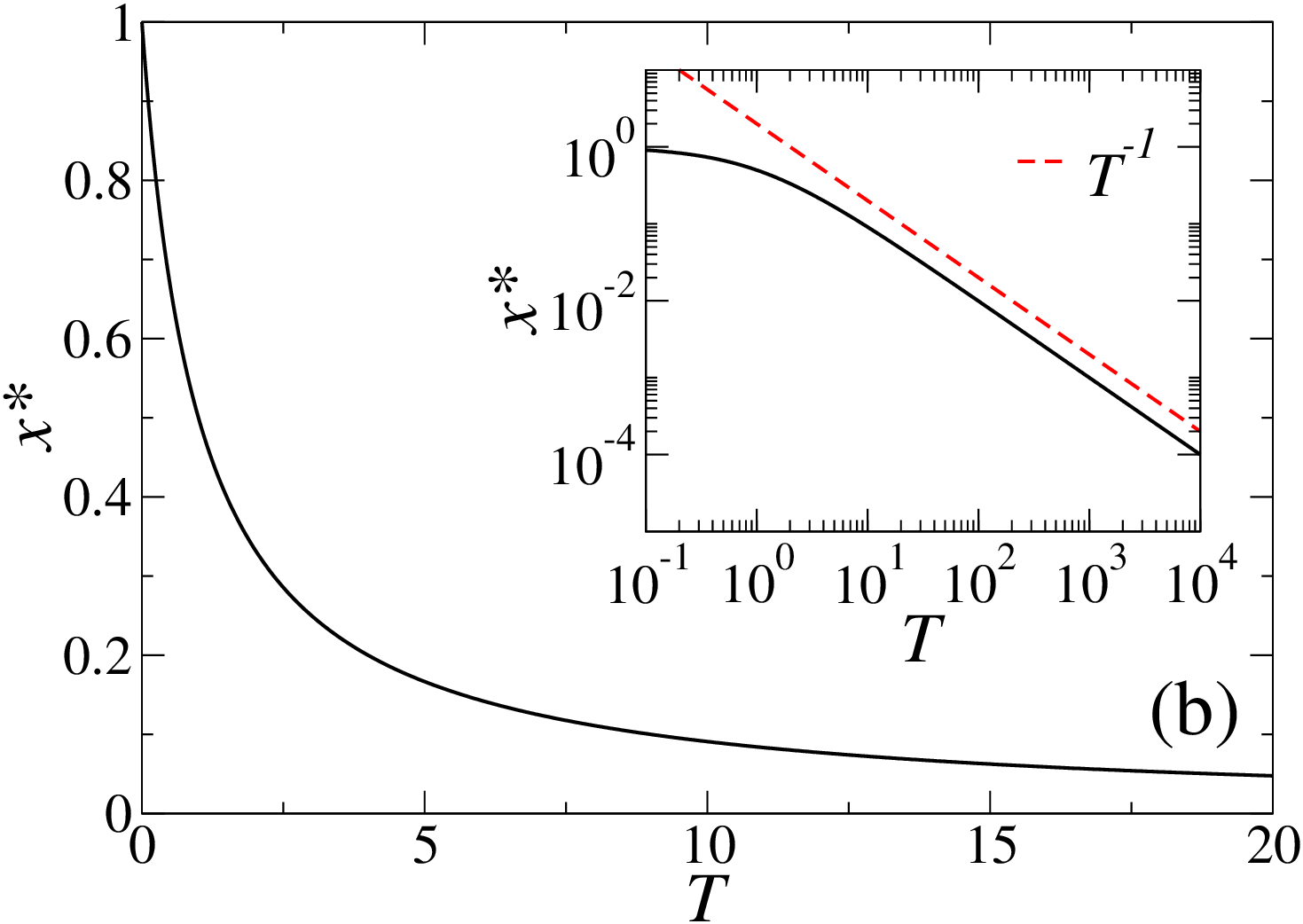}
\end{center}
\caption{(a) Temporal evolution of the population-level sustained attention $x(t)$ for different values of the digital exposure parameter $T$, showing exponential relaxation toward a unique stationary state. (b) Stationary attention level $x^*$ as a function of $T$, illustrating a continuous decrease with increasing exposure. The inset shows the same data in log-log scale, highlighting the asymptotic behavior  $x^*\sim T^{-1}$. The fixed parameters are $r=1.0$ and $\alpha=1.0$.}
\label{fig1}
\end{figure}

Results from the analytical solution of Eq. \eqref{eq1} are shown in Fig. \ref{fig1}. The parameters are $r=1.0, \alpha=1.0$ and the initial condition is $x_0=1.0$. In panel (a), we present the time evolution $x(t)$ of the population-level sustained attention for different values of the intensity of the digital exposure parameter $T$. As previously discussed, the system relaxes toward stationary states $x^*=x(t\to\infty)$, which decrease as $T$ increases. In panel (b), we show the stationary value $x^*$ as a function of $T$, illustrating a continuous decay with increasing exposure. The inset displays the same quantity in the log-log scale, where the asymptotic behavior $x^*\sim T^{-1}$ is clearly observed for large $T$. This behavior is consistent with the analytical expression $x^{*} = r/(r + \alpha\,T)$,  confirming the absence of any critical behavior in the system.

Before introducing a nonlinear extension of the model presented in this subsection, it is useful to further clarify the structural interpretation of the dynamics. The evolution equation \eqref{eq1} can be viewed as the result of two competing effective processes acting on the same macroscopic variable. Rather than representing detailed microscopic mechanisms, these contributions define a coarse-grained balance between restoration and degradation of sustained attention.

The recovery term introduces a stabilizing tendency that drives the system toward high-attention states, while the degradation term acts as a persistent external perturbation that continuously shifts the system away from this regime. In the linear formulation, this competition leads to a single stable stationary state, whose position depends on the intensity of the digital environment.

This interpretation highlights that the collective behavior does not arise from interactions between individuals, but from their shared exposure to a common external field. As a consequence, the model describes a continuous reshaping of the stable cognitive state rather than a transition between competing configurations.

As will be discussed in the next subsection, introducing nonlinear contributions allows one to capture additional features of this degradation process without altering its fundamental interpretation.


\subsection{Nonlinear extension}

A natural extension of the linear model presented in subsection 2.2 is to include a nonlinear contribution to the degradation process. The nonlinear term is introduced directly at the macroscopic level as a phenomenological correction to the degradation process and should not be interpreted as the average of an underlying microscopic quadratic interaction. In this case, Eq. \eqref{eq1} is modified as
\begin{equation} \label{eq4}
\frac{dx}{dt}=r(1-x) - \alpha\,T\,x - \beta\,T\,x^{2} ~.
\end{equation}
\noindent
The first two terms on the right-hand side of Eq. \eqref{eq4} are identical to those in Eq. \eqref{eq1}. The new contribution is the third term, $-\beta\,T\,x^{2}$, which introduces a nonlinear component to the degradation process.

This term accounts for the fact that the erosion of sustained attention may not be purely proportional to its current level. Instead, under persistent exposure to digital stimuli, maintaining coherent cognitive states may become increasingly difficult, leading to a superlinear degradation effect. In this sense, the nonlinear term captures an amplification mechanism associated with high-intensity stimulation.

The quadratic form is not intended to be unique; it represents the simplest analytically tractable nonlinear correction capable of capturing an amplification of the degradation process. More general nonlinear functions could be considered, but the quadratic choice allows us to isolate the effect of nonlinear degradation while keeping the model minimal.

Regarding the nonlinear term $- \beta\,T\,x^{2}$, empirical studies suggest that sustained exposure to digitally mediated environments is associated with cumulative impairments in cognitive control and attentional filtering, particularly in individuals engaged in high levels of media multitasking \cite{gazzaley,ophir}. These findings indicate that the degradation of attention may not be purely proportional to its current level, but instead may reflect an increasing difficulty in maintaining coherent cognitive states under persistent stimulation. Within this perspective, the quadratic term provides a minimal phenomenological representation of such nonlinear effects, capturing the idea that attention degradation can accelerate as exposure increases.

From a mathematical perspective, Eq. \eqref{eq4} belongs to the class of Riccati equations and bears a formal resemblance to the Verhulst logistic equation \cite{boyce}, although the interpretation of the dynamical variable and parameters is entirely different.

Taking into account the dimensionless variables $\tilde{t}=r\,t$ and $\theta_1=\alpha\,T/r$ introduced in section 2.2, and introducing a third variable $\theta_2=\beta\,T/r$, Eq. \eqref{eq4} becomes
\begin{equation} \label{eq4_2}
\frac{dx}{d\tilde{t}}=(1-x) - \theta_1\,x - \theta_2\,x^{2} ~.
\end{equation}
\noindent
showing that the nonlinear dynamics is governed by two independent dimensionless control parameters, $\theta_1$ and $\theta_2$, associated with the linear and nonlinear degradation mechanisms, respectively.

The stationary solution $x^*$  is obtained by imposing $dx/dt = 0$ in Eq. \eqref{eq4}, which leads to $\beta\,T\,x^{2} + (r + \alpha\,T)\,x - r = 0$. Solving the quadratic equation, we obtain
\begin{equation} \label{eq5}
x^* = \frac{-(r+\alpha\,T) + \sqrt{(r+\alpha\,T)^2 + 4\beta\,T\,r}}{2\beta\,T} ~,
\end{equation}
\noindent
where the positive branch is chosen to ensure $x^* \in [0,1]$. In terms of the dimensionless variables, Eq. \eqref{eq5} can be rewritten as
\begin{equation} \label{eq5_2}
x^* = \frac{-(1+\theta_1) + \sqrt{(1+\theta_1)^2 + 4\theta_2}}{2\theta_2} ~.
\end{equation}
The full-dependent analytical solution $x(t)$ of the nonlinear model is presented in Appendix A.

\begin{figure}[t]
\begin{center}
\vspace{6mm}
\includegraphics[width=0.35\textwidth,angle=0]{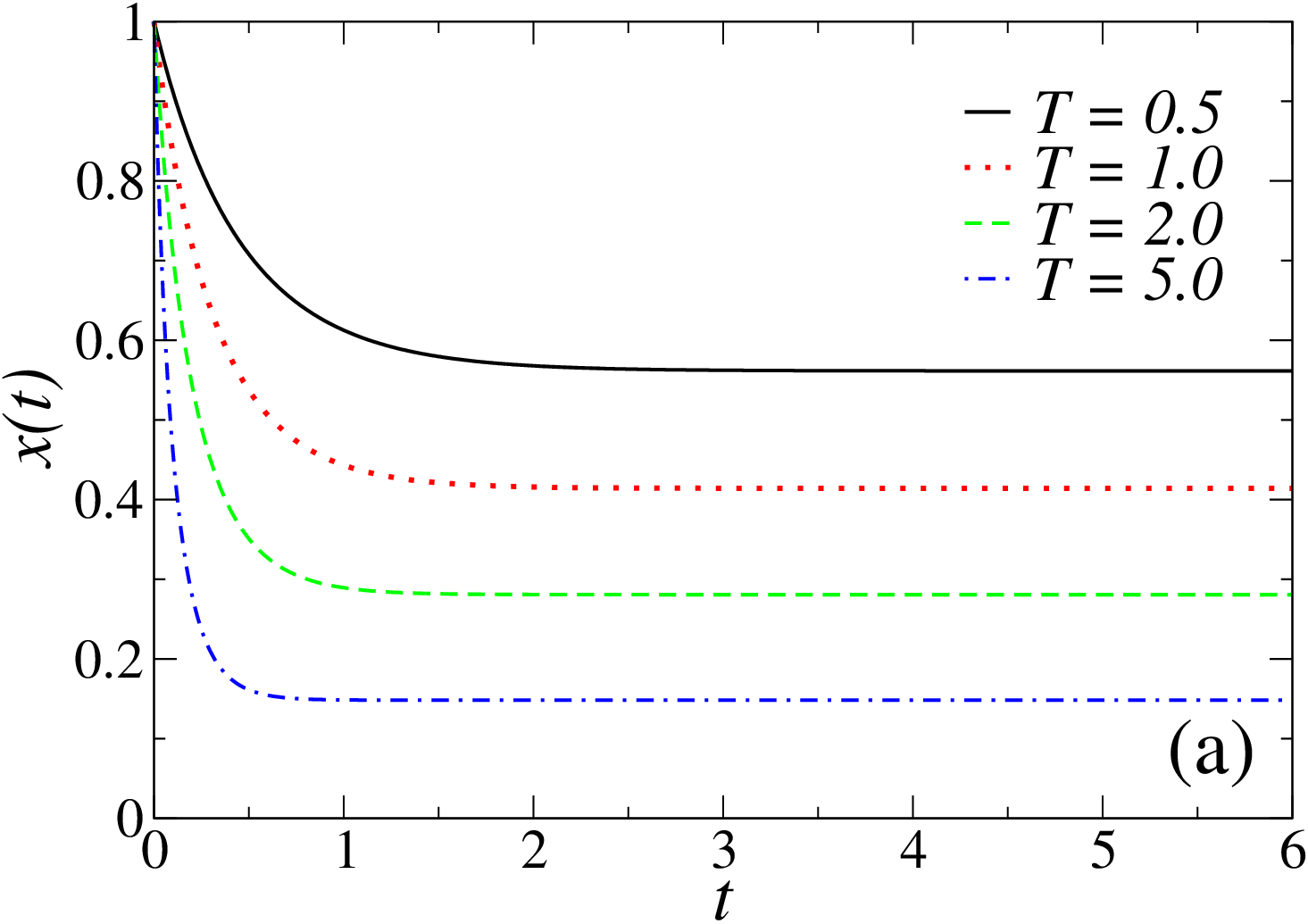}
\hspace{0.3cm}
\includegraphics[width=0.35\textwidth,angle=0]{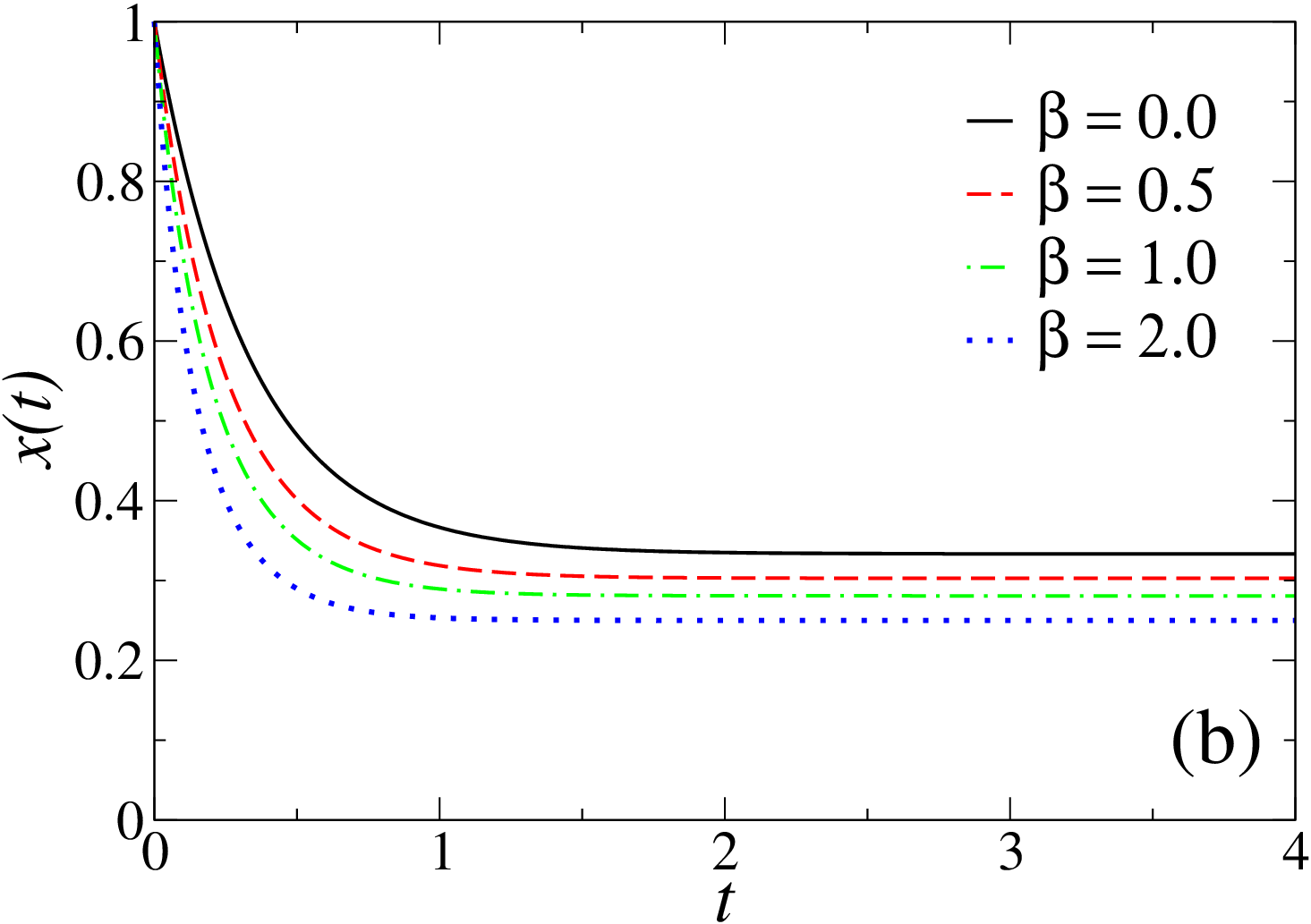}
\end{center}
\caption{(a) Temporal evolution of the population-level sustained attention $x(t)$ for different values of the digital exposure parameter $T$, for a fixed nonlinear coefficient $\beta=1.0$, showing relaxation toward a unique stationary state. (b) Temporal evolution of $x(t)$ for different values of $\beta$, at fixed $T=2.0$, highlighting the role of nonlinear degradation. Larger values of $\beta$ lead to a stronger suppression of the stationary attention level and a faster convergence toward stationary states. The other parameters are $r=1.0$ and $\alpha=1.0$.}
\label{fig2}
\end{figure}


Equation \eqref{eq5_2} makes explicit that the stationary attention level is controlled by the two dimensionless parameters $\theta_1$ and $\theta_2$. As in the linear case, increasing digital exposure leads to lower values of $x^{*}$. However, the nonlinear contribution modifies the functional dependence of the stationary solution on the exposure intensity, leading to a more pronounced reduction of attention for large values of stimulation.

In particular, the nonlinear contribution introduces a curvature in the dependence of $x^*$ on $T$ (or $\theta_2$), which reflects an accelerated degradation regime. This behavior is consistent with the interpretation that sustained exposure to highly stimulating environments may produce cumulative effects on cognitive performance.

In Fig. \ref{fig2}, we show the temporal evolution $x(t)$ of the population-level sustained attention for different values of the digital exposure $T$, for a fixed nonlinear coefficient $\beta=1.0$. The results were obtained through numerical integration of Eq. \eqref{eq4} using a fourth-order Runge-Kutta method. The curves relax towards stationary states, with a faster convergence compared to the linear model presented in subsection 2.2. Panel (b) of Fig. \ref{fig2} illustrates the effect of the nonlinear coefficient $\beta$ on the temporal evolution $x(t)$, for fixed $T=2.0$. Increasing $\beta$ leads to a more pronounced degradation and a faster convergence toward lower stationary attention levels.

\begin{figure}[t]
\begin{center}
\vspace{6mm}
\includegraphics[width=0.35\textwidth,angle=0]{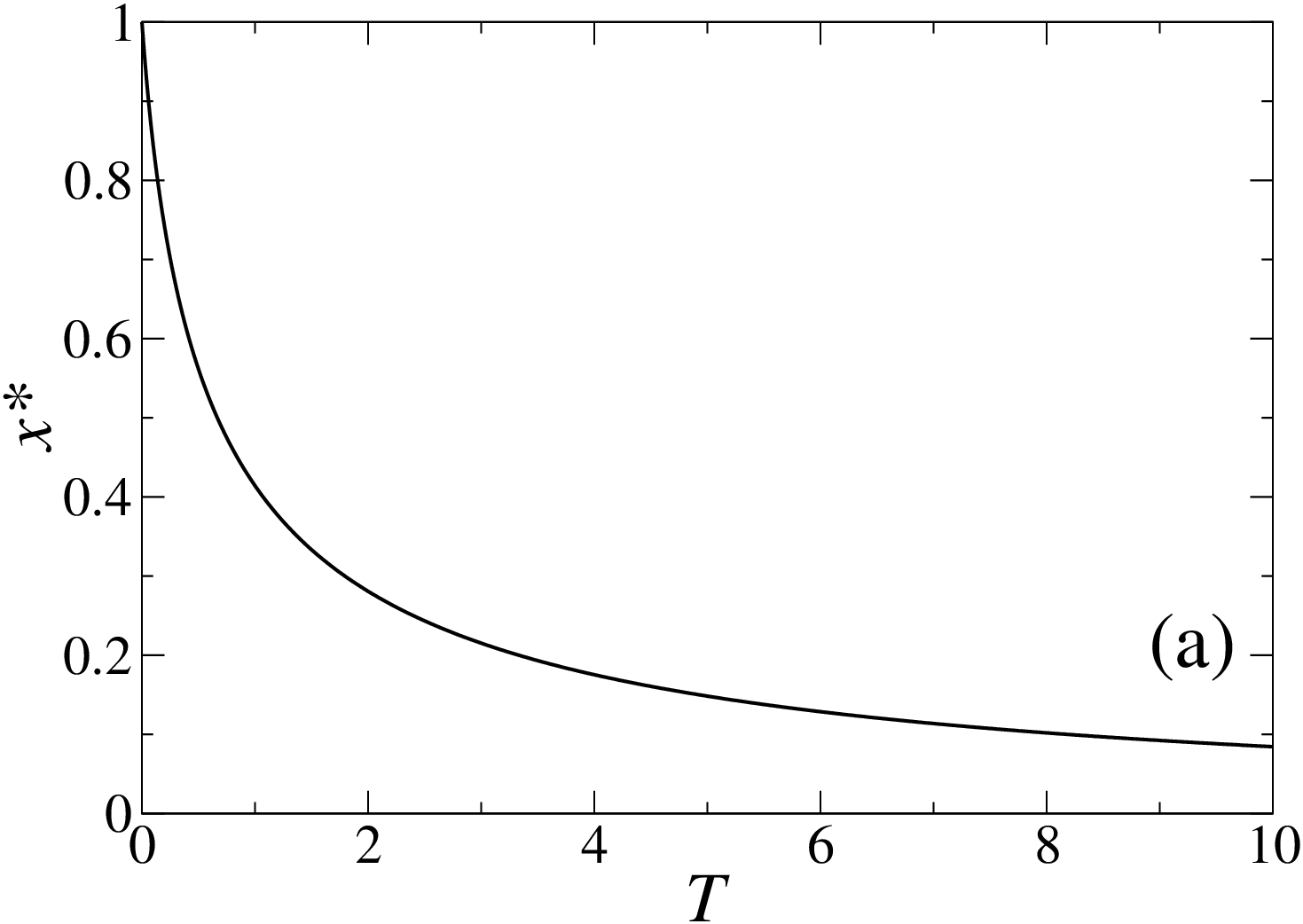}
\hspace{0.3cm}
\includegraphics[width=0.35\textwidth,angle=0]{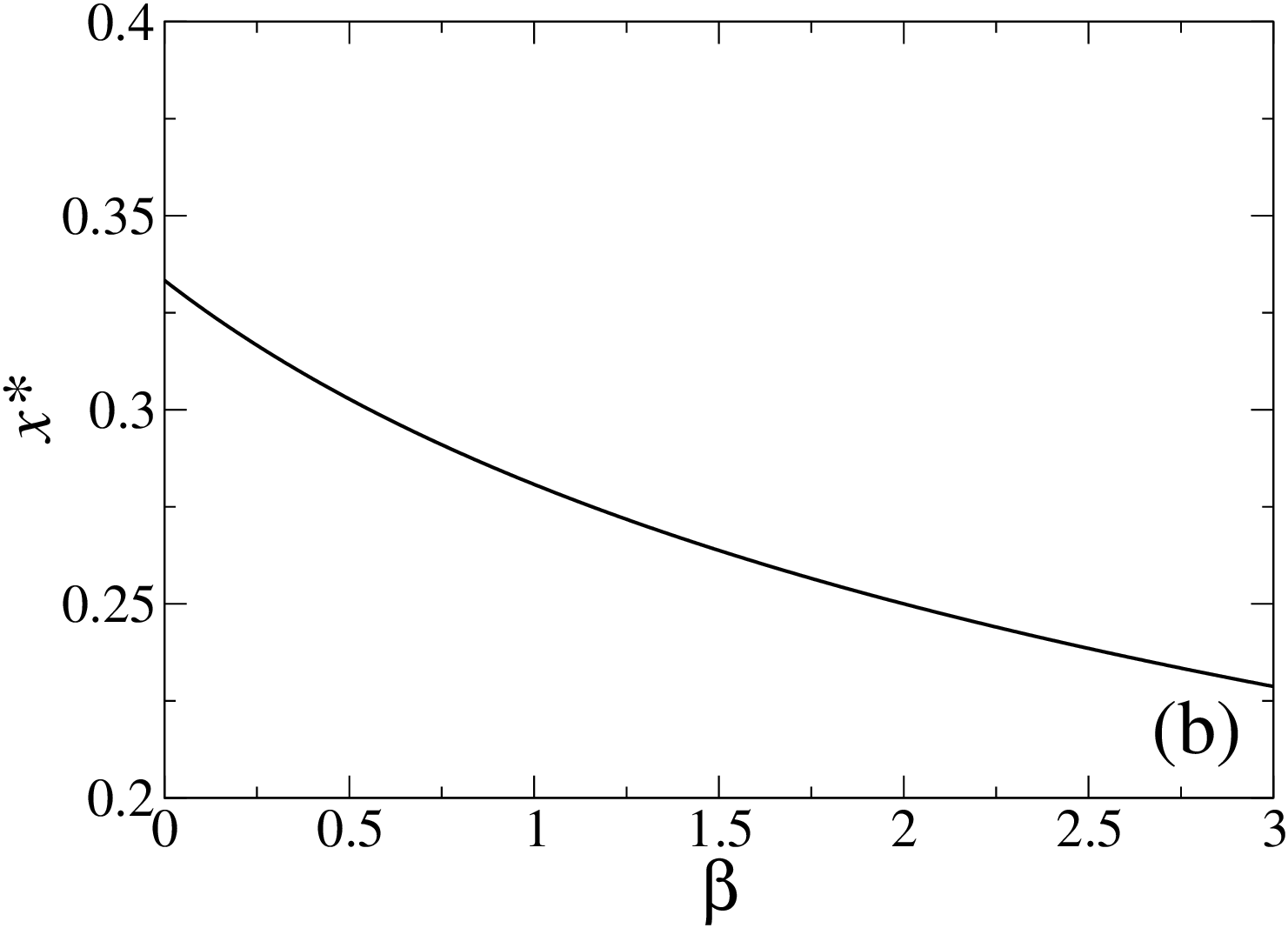}
\end{center}
\caption{(a) Stationary population-level sustained attention $x^*$ as a function of the digital exposure parameter $T$, for a fixed nonlinear coefficient $\beta=1.0$, showing a continuous decrease with increasing exposure. (b) Stationary attention level $x^*$ as a function of $\beta$, for fixed $T=2.0$, highlighting the role of nonlinear degradation. Larger values of $\beta$ lead to a stronger suppression of the stationary attention level. The other parameters are $r=1.0$ and $\alpha=1.0$.}
\label{fig3}
\end{figure}

In Fig. \ref{fig3}, we show the analytical stationary values $x^*$, obtained from Eq. \eqref{eq5}. The parameters are $r=1.0, \alpha=1.0$ and the initial condition is $x_0=1.0$. In panel (a), $x^*$ is shown as a function of the exposure parameter $T$, for a fixed $\beta=1.0$. We observe that increasing $T$ leads to a continuous decrease in the stationary attention level, indicating a progressive degradation of sustained attention as digital exposure becomes more intense. In panel (b), $x^*$ is plotted as a function of $\beta$, for fixed $T=2.0$. The linear model is recovered for $\beta=0.0$, for which one obtains $x^*=1/3$, in agreement with Eq. \eqref{eq2}. Increasing $\beta$ leads to a stronger suppression of the stationary attention level, highlighting the role of nonlinear degradation in amplifying the loss of sustained attention.

To further characterize the collective dynamics, it is useful to introduce an effective potential description. The evolution equation
\begin{equation}
\frac{dx}{dt}=r(1-x)-\alpha T x-\beta T x^2
\end{equation}
\noindent
can be written in gradient form as $\frac{dx}{dt}=-\frac{dV}{dx}$, where $V(x)$ is an effective potential governing the macroscopic dynamics of the system. By direct integration, one obtains
\begin{equation}
V(x)= -r x+\frac{1}{2}(r+\alpha T)x^2+\frac{1}{3}\beta T x^3,
\end{equation}
\noindent
up to an irrelevant additive constant. It should be noted that the effective potential is physically meaningful only within the interval $0 \leq x \leq 1$, which corresponds to the admissible range of the collective attention variable. Therefore, the behavior of $V(x)$ outside this domain has no physical interpretation in the context of the present model.

Within this formulation, the collective attention level $x$ evolves as a relaxational variable in an asymmetric potential landscape. The different contributions to $V(x)$ have clear interpretations:
\begin{itemize}

\item The linear term $-r\,x$ encodes the tendency toward recovery of sustained attention through restorative processes such as rest and offline cognitive activities.

\item The quadratic term $\frac{1}{2}(r+\alpha\,T)x^2$ provides stabilization and incorporates both intrinsic relaxation and linear degradation induced by the digital environment.

\item The cubic term $\frac{1}{3}\beta\,T\,x^3$ introduces an intrinsic asymmetry that biases the system toward lower attention levels as the intensity of digital exposure increases.

\end{itemize}

The stationary state of the system corresponds to the minimum of the effective potential. As the parameter $T$, representing the intensity of the digital environment, increases, the shape of $V(x)$ is progressively deformed.

In particular: (i) for small $T$, the potential minimum is located at high values of $x$, corresponding to a regime of sustained attention; (ii) as $T$ increases, the minimum continuously shifts toward lower values of $x$, indicating a degradation of the collective attention level. This behavior reflects a smooth displacement of the stable macroscopic state rather than the emergence of competing equilibria.

\begin{figure}[t]
\begin{center}
\vspace{6mm}
\includegraphics[width=0.35\textwidth,angle=0]{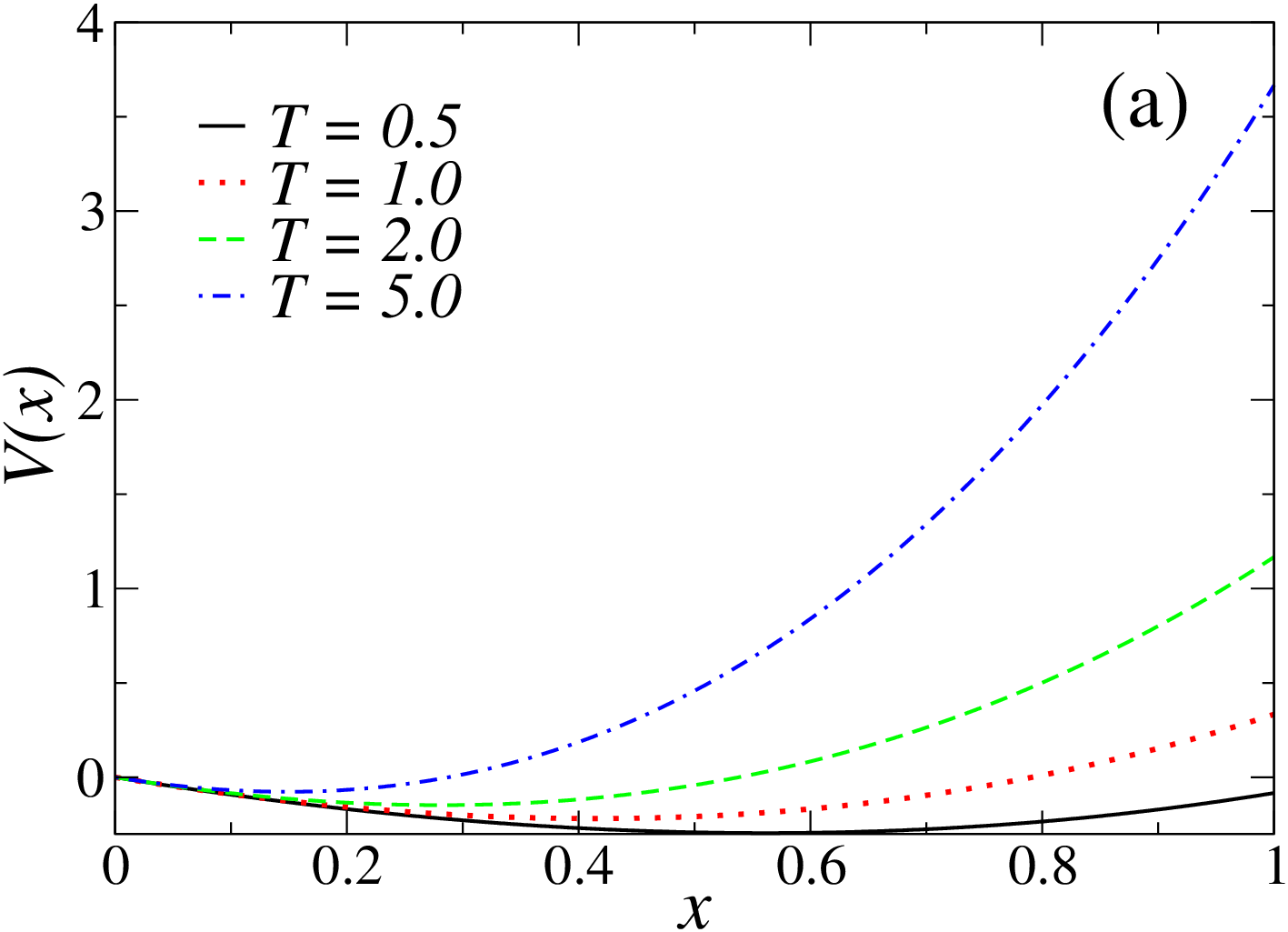}
\hspace{0.3cm}
\includegraphics[width=0.35\textwidth,angle=0]{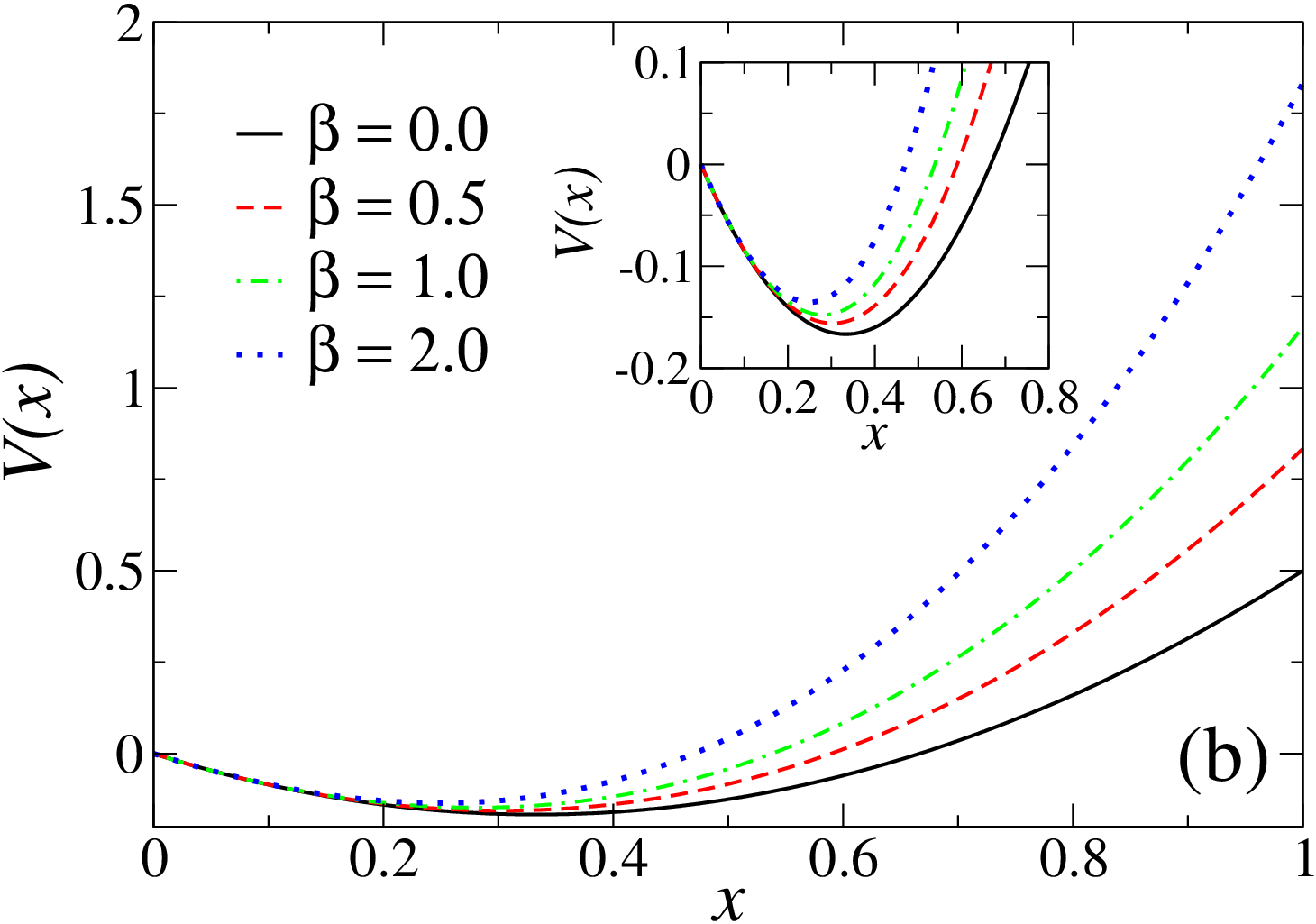}
\end{center}
\caption{(a) Effective potential $V(x)$ for different values of the digital exposure parameter $T$, for fixed $\beta=1.0$, illustrating a continuous deformation of the landscape and a shift of the minimum toward lower values of $x$. (b) Effective potential $V(x)$ for different values of $\beta$, at fixed $T=2.0$, highlighting the effect of nonlinear degradation. The inset shows a zoom near the minimum, clearly illustrating the shift of the equilibrium position as $\beta$ increases. In all cases, the potential remains single-welled, indicating the absence of bistability. The fixed parameters are $r=1.0$ and $\alpha=1.0$.}
\label{fig4}
\end{figure}

Figure \ref{fig4} shows the effective potential associated with the nonlinear model. Panel (a) presents $V(x)$ as a function of $x$ for different values of $T$, at fixed $\beta=1.0$, while panel (b) shows $V(x)$ as a function of $x$ for different values of $\beta$, at fixed $T=2.0$. As the parameters vary, the potential is continuously deformed while retaining a single minimum, which shifts toward lower values of $x$. This behavior reflects the progressive suppression of sustained attention induced by increasing digital exposure and nonlinear effects. The inset in panel (b) highlights the region near the minimum, clearly illustrating its displacement without the emergence of additional stable states.

In contrast with models of social contagion or opinion dynamics, where multiple stable states may coexist, the present framework is intrinsically asymmetric and admits a single stable equilibrium.

Therefore, the key phenomenon described here is not a discontinuous transition between competing collective states, but rather a continuous reshaping of the stable cognitive regime under the influence of an external environment.

This perspective suggests that the impact of digital overstimulation should be understood as a gradual deformation of the dynamical landscape governing collective attention. As the intensity of exposure increases, the system is progressively driven toward regimes of reduced sustained attention, without requiring abrupt transitions or bistability.


\section{Empirical grounding of the dynamical mechanisms}

A central assumption of the present framework is that the collective dynamics of sustained attention can be understood as the result of competing recovery and degradation processes. While the model is formulated at a macroscopic level, both mechanisms are supported by a growing body of empirical evidence.

\vspace{0.4cm}

\textbf{i) Evidence for intrinsic recovery of attention:}
\\

The recovery term reflects the well-established idea that attentional resources are not static, but can be restored through appropriate conditions. Research in cognitive psychology and neuroscience has consistently shown that sustained attention is a limited resource subject to depletion and recovery \cite{kahneman,posner}.

More recent studies indicate that uninterrupted cognitive engagement, reduced task switching, and rest periods contribute to restoring attentional capacity. Experimental work has shown that even brief interruptions can significantly impair performance in tasks requiring sustained focus, while periods of low stimulation or structured activity can partially restore cognitive efficiency \cite{leroy}.

In this context, the recovery term $r(1-x)$ should be interpreted as an effective representation of these restorative processes, rather than as a detailed mechanistic description. This interpretation is consistent with the notion that cognitive systems exhibit intrinsic relaxation dynamics toward baseline functional states, in analogy with relaxational processes in driven systems \cite{Hohenberg}.

\vspace{0.5cm}

\noindent
\textbf{ii) Evidence for degradation under digital stimulation:}
\\

A large and growing body of literature has documented the impact of digital environments on attentional processes. One of the most consistent findings is the increase in task switching and the reduction of sustained focus in screen-mediated contexts.

Empirical studies of human-computer interaction have shown that individuals frequently switch between tasks and windows when interacting with digital devices, often at timescales of seconds to minutes \cite{mark1,mark2,mark3}. These patterns are associated with reduced depth of processing and increased cognitive fragmentation.

In parallel, experimental and observational studies suggest that frequent exposure to fast-paced, reward-driven digital content may alter attentional habits over time \cite{gazzaley}. The constant availability of novel stimuli creates an environment in which sustained attention competes with rapidly changing inputs, potentially biasing behavior toward shorter attention cycles.

At a broader scale, population-level analyses have reported correlations between increased screen time and indicators of reduced well-being, increased distractibility and difficulties in maintaining prolonged cognitive engagement \cite{Twenge}. Within this framework, the degradation term can be interpreted as an effective representation of these cumulative and environment-driven attentional shifts.

\vspace{0.5cm}

\noindent
\textbf{iii) Nonlinear effects and amplification mechanism:}
\\

Beyond direct effects, there is also evidence suggesting that the impact of digital environments may be nonlinear. For instance, heavy multitasking and frequent task switching have been associated with reduced cognitive control and increased susceptibility to distraction, even outside digital contexts \cite{gazzaley,ophir}.

These findings suggest the presence of feedback mechanisms in which sustained exposure to fragmented stimuli may progressively alter the ability to maintain attention itself. In this sense, attention degradation may not be simply proportional to its current level, but instead may be amplified under persistent high-intensity stimulation.

Such effects are consistent with the inclusion of a nonlinear degradation term in the model, which provides a minimal phenomenological representation of this amplification mechanism. This interpretation aligns with the notion that complex systems under sustained external forcing may exhibit nonlinear response functions.

\vspace{0.5cm}

\noindent
\textbf{iv) From individual behavior to collective dynamics:}
\\

Although most empirical studies focus on individual cognition, their implications naturally extend to the population level. Digital technologies create a shared environment characterized by similar patterns of stimulation, interaction and information flow.

In this sense, the collective nature of the phenomenon does not require explicit interactions between individuals. Instead, it emerges from their common exposure to the same external conditions. This view is further supported by empirical and modeling studies showing that attention patterns in digital environments emerge from the interplay between individual behavior and shared external conditions, often leading to collective dynamics shaped by common exposure and feedback mechanisms \cite{wagner,perra}.

This justifies modeling the population-level sustained attention as a macroscopic variable subject to a global control parameter $T$. This perspective is consistent with the description of driven systems in statistical physics, where collective behavior arises from a common external field rather than from direct interactions.

\vspace{0.5cm}

Taken together, these empirical findings support the key assumptions of the model: (i) attention is a dynamic resource subject to recovery, (ii) digital environments introduce persistent perturbations that degrade sustained attention, and (iii) these effects may be nonlinear and cumulative.

Within this context, the proposed dynamical framework provides a minimal and analytically tractable description that connects individual-level observations with collective macroscopic behavior. This connection reinforces the relevance of a coarse-grained approach to modeling attention dynamics in modern digital environments.


\section{Final Remarks}   

In this work, we have proposed a minimal dynamical framework to describe the collective evolution of sustained attention under continuous exposure to digital environments. By introducing a macroscopic variable representing the population-level sustained attention of the population, we model the dynamics as the result of a competition between intrinsic cognitive recovery and degradation induced by digital stimulation. The present framework should therefore be interpreted as a phenomenological coarse-grained description formulated directly at the population level, complementary to microscopic or agent-based approaches commonly employed in complex systems research.

The linear formulation provides a baseline description in which the stationary attention level decreases monotonically with increasing exposure, while remaining analytically tractable. The nonlinear extension captures additional features of the degradation process by introducing an amplification mechanism that leads to a more pronounced reduction of attention under high-intensity stimulation. In both cases, the dynamics can be interpreted as a relaxational process in an effective potential landscape, whose minimum is continuously shifted by the external environment. This provides a unified physical picture of how digital exposure reshapes collective attention dynamics.

Importantly, the model does not rely on interaction-driven mechanisms or on the emergence of competing stable states. Instead, it describes a continuous displacement of the stable cognitive regime under the influence of a persistent external perturbation. This perspective suggests that large-scale changes in attentional behavior may arise from sustained environmental forcing, rather than from abrupt transitions or interaction-driven criticality. This highlights an alternative route to collective behavior in cognitive systems.

Empirical studies consistently report moderate but persistent effects of digital media consumption on sustained attention and inhibitory control, supporting a scenario in which cognitive performance degrades progressively with exposure rather than through abrupt transitions. For example, recent studies have reported a significant positive association between media multitasking and attention deficits, indicating that higher levels of multitasking are linked to poorer attentional performance \cite{chen}. Other empirical works predominantly point to gradual changes in attentional behavior rather than abrupt collective shifts \cite{Nguyen,Rioja}. In this context, the minimal formulation adopted here appears sufficient to capture the essential features of the phenomenon, as it naturally reproduces a continuous degradation of attention with increasing exposure.

A natural question in the context of statistical physics is whether the system exhibits a phase transition or bistability, as commonly observed in models of collective behavior such as opinion dynamics or social contagion \cite{rmp}. In the present framework, however, the effective potential remains single-welled for all parameter values considered, and the system displays a unique stable equilibrium.

Rather than representing a limitation, this feature reflects a fundamental aspect of the phenomenon being modeled. The dynamics considered here does not arise from competition between equivalent collective states or interaction-driven symmetry breaking. Instead, it is governed by the interplay between intrinsic recovery mechanisms and a persistent external perturbation associated with digital exposure. This interpretation is consistent with empirical evidence pointing to gradual changes in attentional behavior, without indications of abrupt collective transitions \cite{chen,Nguyen,Rioja}.

From a physical standpoint, the dynamics can be interpreted as a relaxational process in a deformable potential landscape. In this picture, the digital environment continuously modifies the shape of the effective potential, shifting its minimum without generating additional stable states. As a result, the system is progressively biased toward lower values of sustained attention.

This framework suggests that digital exposure acts as a persistent source of perturbations that continuously destabilizes coherent cognitive states, rather than as a mechanism leading to the emergence of competing collective configurations.

It is worth noting that additional nonlinear terms could be introduced to generate multiple equilibria and phase transitions. However, such extensions would increase the number of free parameters and reduce the interpretability of the model, without clear empirical support for this additional complexity.

Given that available empirical evidence predominantly points to gradual changes in attentional behavior rather than abrupt collective shifts \cite{mark1,mark2,gazzaley,ophir}, the minimal formulation adopted here appears sufficient to capture the essential features of the phenomenon.

The present results suggest that the impact of digital environments on collective attention does not require critical thresholds or abrupt transitions to be significant. Instead, even a continuous displacement of the stable cognitive regime can lead to substantial macroscopic effects as the intensity of exposure increases. This highlights a mechanism through which gradual changes at the microscopic level can translate into significant collective outcomes.

The simplicity of the model is intentional, as it allows us to isolate the essential mechanism underlying the observed collective behavior. In this sense, the present framework should be viewed as a first step toward a coarse-grained description of attention dynamics at the population level.

The present framework assumes a homogeneous population characterized by effective average parameters. In reality, individuals differ in their recovery rates, sensitivities to digital stimulation, and patterns of media consumption. Moreover, digital interactions occur on heterogeneous networks whose structure may influence the collective dynamics of attention. While such effects are neglected in the present minimal formulation, incorporating population heterogeneity and network-mediated interactions constitutes a natural extension of the model.

Another important limitation of the present framework is the assumption that the digital exposure parameter $T$ acts as an external and time-independent control variable. In contemporary digital platforms, however, exposure levels are often shaped by recommendation algorithms that continuously adapt content delivery according to user engagement and behavioral responses. As a result, the intensity of digital stimulation may itself evolve dynamically in response to the collective attention state of the population, generating a coupled human-algorithm feedback system. While incorporating such feedback mechanisms lies beyond the scope of the present minimal model, exploring co-evolutionary dynamics between collective attention and algorithmic optimization represents an important direction for future research.

Future work may also include quantitative calibration of the model using empirical measures of digital exposure, multitasking behavior and sustained attention performance. Incorporating memory effects, heterogeneous populations or time-dependent exposure may also provide a more realistic description of the dynamics associated with digital media exposure. In addition, more direct quantitative comparisons between model predictions and empirical measures of digital exposure and attention may provide further insight into the mechanisms underlying the observed behavior. Another possible extension would be the derivation of the proposed macroscopic equations from explicit microscopic or agent-based attention dynamics. These developments may contribute to a more comprehensive understanding of how digital environments shape cognitive processes in modern societies.


\appendix

\section{Full analytical solution of the nonlinear model}

Starting from the dimensionless nonlinear model, Eq. \eqref{eq4_2},
\begin{equation}
\frac{dx}{d\tilde{t}} = 1-(1+\theta_1)x-\theta_2 x^2,
\end{equation}
\noindent
the stationary solutions are obtained from
\begin{equation}
1-(1+\theta_1)x-\theta_2 x^2=0,
\end{equation}
\noindent
yielding
\begin{equation}
x_{\pm} = \frac{-(1+\theta_1)\pm\sqrt{(1+\theta_1)^2+4\theta_2}}{2\theta_2}.
\end{equation}
\noindent
as discussed in the text. The differential equation can then be written as
\begin{equation}
\frac{dx}{d\tilde{t}} = -\theta_2(x-x_+)(x-x_-).
\end{equation}

Integrating by partial fractions, one obtains
\begin{equation}
x(\tilde{t}) = \frac{x_+-Kx_-e^{-\Lambda\tilde{t}}}{1-Ke^{-\Lambda\tilde{t}}},
\end{equation}
\noindent
where
\begin{eqnarray}
K & = & \frac{x_0-x_+}{x_0-x_-} \\
\Lambda & = & \sqrt{(1+\theta_1)^2+4\theta_2}.
\end{eqnarray}
\noindent
and $x_0=x(0)$. The physical stationary state $x^{*}$ corresponds to $x_+$, whereas $x_{-}$ is always negative. In the limit $\theta_2\to 0$ (equivalent to $\beta\to 0$), the solution reduces continuously to the exponential relaxation obtained for the linear model, namely Eq. \eqref{eq3_2} of the text.


\section*{Acknowledgments}

The author acknowledges partial financial support from the Brazilian scientific funding agency Conselho Nacional de Desenvolvimento Cient\'ifico e Tecnol\'ogico  (CNPq, Grants 308643/2023-2 and 406820/2025-2).

\bibliographystyle{elsarticle-num-names}

\begin{thebibliography}{00}


\bibitem{arruda}
H. F. de Arruda, Y. Moreno, \textit{The social consequences of AI delegation},   arXiv:2606.11058 (2026).




\bibitem{yakura}  
H. Yakura et. al., \textit{Empirical evidence of Large Language Model's influence on human spoken communication}, arXiv:2409.01754 (2024).


  


\bibitem{Twenge}
J. M. Twenge, \textit{iGen: Why Today's Super-Connected Kids Are Growing Up Less Rebellious, More Tolerant, Less Happy--and Completely Unprepared for Adulthood--and What That Means for the Rest of Us} (Atria Books, 2017).


\bibitem{mark1}
G. Mark, D. Gudith, U. Klocke, \textit{The Cost of Interrupted Work: More Speed and Stress}, Proceedings of the SIGCHI Conference on Human Factors in Computing Systems, 2008, pp. 107–110.
  

\bibitem{mark2}
G. Mark, \textit{Attention Span: A Groundbreaking Way to Restore Balance, Happiness and Productivity} (Hanover Square Press, 2023).


\bibitem{mark3}
G. Mark, S. T. Iqbal, M. Czerwinski, P. Johns, \textit{Bored mondays and focused afternoons: the rhythm of attention and online activity in the workplace}, CHI '14: Proceedings of the SIGCHI Conference on Human Factors in Computing Systems
Pages 3025-3034 (2014).

  

\bibitem{gazzaley}
A. Gazzaley, L. D. Rosen, \textit{The Distracted Mind: Ancient Brains in a High-Tech World} (MIT Press, Cambridge, MA, 2016).



\bibitem{ophir}
E. Ophir et. al, \textit{Cognitive control in media multitaskers}, Proc. Natl. Acad. Sci. USA 106 (37), 15583-15587 (2009).


\bibitem{leroy}
S. Leroy, \textit{Why is it so hard to do my work? The challenge of attention residue when switching between work tasks}, Organizational Behavior and Human Decision Processes 109(2), 168-181 (2009).


\bibitem{kahneman}
D. Kahneman, \textit{Attention and Effort} (Prentice Hall, 1973).



\bibitem{Traulsen2005}
A. Traulsen, J. C. Claussen, C. Hauert, \textit{Coevolutionary dynamics: from finite to infinite populations}, Phys. Rev. Lett. 95, 238701 (2005).


\bibitem{Traulsen2006}
A. Traulsen, J. C. Claussen, C. Hauert, \textit{Coevolutionary dynamics in large, but finite populations},  Phys. Rev. E 74, 011901 (2006).


  
\bibitem{Tu2023}
C. Tu, P. D'Odorico, Z. Li, S. Suweis, \textit{The emergence of cooperation from shared goals in the governance of common-pool resources}, Nature Sustainability 6, 139-147 (2023).



\bibitem{rmp}
C. Castellano, S. Fortunato, V. Loreto, \textit{Statistical physics of social dynamics}, Rev. Mod. Phys. 81, 591 (2009).  


\bibitem{vacaruiz}
C. V. Ruiz, L. M. Aiello, A. Jaimes, \textit{Modeling dynamics of attention in social media with user efficiency}, EPJ Data Sci. 3, 5 (2014).  


\bibitem{ojer}
J. Ojer et. al., \textit{Modeling individual attention dynamics on online social media}, arXiv:2507.01511 (2026).


\bibitem{galam}
Sociophysics: A review of Galam models, Int. J. Mod. Phys. C 19, 409-440 (2008).


\bibitem{racism}
N. Crokidakis, L. Sigaud, \textit{Nonequilibrium phase transitions in a racism-spreading model with interaction-driven dynamics}, Eur. Phys. J. B 99, 34 (2026).




\bibitem{Hohenberg}
P. C. Hohenberg, B. I. Halperin, \textit{Theory of dynamic critical phenomena}, Rev. Mod. Phys 49, 435 (1977).


\bibitem{Goldenfeld}
N. Goldenfeld, \textit{Lectures On Phase Transitions And The Renormalization Group } (CRS Press, 2019).


\bibitem{boyce}
W. E. Boyce, R. C. DiPrima, D. B. Meade, \textit{Elementary differential equations and boundary value problems} (John Wiley $\&$ Sons, 2021).

  

\bibitem{posner}
 M. I. Posner, S. E. Petersen, \textit{The attention system of the human brain}, Annu. Rev. Neurosci. 13, 25-42 (1990). 


\bibitem{wagner}
C. Wagner et. al., \textit{Ignorance Isn't Bliss: An Empirical Analysis of Attention Patterns in Online Communities}, In: 2012 International Conference on Privacy, Security, Risk and Trust and 2012 International Confernece on Social Computing, $doi:10.1109/SocialCom-PASSAT.2012.33$.


\bibitem{perra}
N. Perra, D. Balcan, B. Gonçalves, A.  Vespignani, \textit{Towards a Characterization of Behavior-Disease Models}, PLoS ONE 6(8): e23084 (2011).

  
\bibitem{chen}
H. Chen et. al., \textit{The relationship between media multitasking and attention: a three-level meta-analysis}, Current Psychology 44, 6326-6347 (2025).


\bibitem{Nguyen}
L. Nguyen et. al., \textit{Feeds, feelings, and focus: A systematic review and meta-analysis examining the cognitive and mental health correlates of short-form video use}, Psychol. Bull. 151(9):1125-1146 (2025).



\bibitem{Rioja}
K. Rioja, S. Cekic, D. Bavelier, S. E. Baumgartner, \textit{Unraveling the Link Between Media Multitasking and Attention Across Three Samples}, Technology, Mind, and Behavior 4(2), 134–150 (2023).





\end{thebibliography}

\end{document}